\documentclass{iau} 
\usepackage{graphicx}
\usepackage[breaklinks=true]{hyperref}                                           
\hypersetup{                                                                     
    bookmarksopen=false,                                                         
    bookmarksnumbered=false,                                                     
    unicode=true,           
    pdftoolbar=true,        
    pdfmenubar=true,        
    pdffitwindow=false,     
    pdfstartview={FitH},    
    pdfnewwindow=true,      
    colorlinks=true,       
    linkcolor=red,          
    citecolor=magenta,        
    filecolor=green,      
    urlcolor=blue           
}                                                                                
\usepackage{threeparttable}

\title[What can Rust do for astrophysics?] 
{What can the programming language\\ Rust do for astrophysics?}

\author[Sergi Blanco-Cuaresma \& Emeline Bolmont]   
{Sergi Blanco-Cuaresma$^1$
 \and Emeline Bolmont$^2$}

\affiliation{$^1$Observatoire de Gen\`eve, Universit\'e de Gen\`eve, \\ CH-1290 Versoix, Switzerland. \\ email: {\tt Sergi.Blanco@unige.ch} \\[\affilskip]
$^2$ NaXys, Department of Mathematics, University of Namur, \\ 8 Rempart de la Vierge, 5000 Namur, Belgium. \\ email: {\tt emeline.bolmont@unamur.be}}

\pubyear{2017}
\volume{325}  
\jname{Astroinformatics}
\editors{M. Brescia, S. G. Djorgovski, E. Feigelson, G. Longo \& S. Cavuoti.}
\begin{document}

\maketitle

\begin{abstract}
The astrophysics community uses different tools for computational tasks such as complex systems simulations, radiative transfer calculations or big data. Programming languages like Fortran, C or C++ are commonly present in these tools and, generally, the language choice was made based on the need for performance.  However, this comes at a cost: safety. For instance, a common source of error is the access to invalid memory regions, which produces random execution behaviors and affects the scientific interpretation of the results.

In 2015, Mozilla Research released the first stable version of a new programming language named Rust. Many features make this new language attractive for the scientific community, it is open source and it guarantees memory safety while offering zero-cost abstraction.

We explore the advantages and drawbacks of Rust for astrophysics by re-implementing the fundamental parts of Mercury-T, a Fortran code that simulates the dynamical and tidal evolution of multi-planet systems.
\end{abstract}

\firstsection 
\section{Introduction}

\begin{figure}[htbp]
\begin{center}
 \includegraphics[trim=150 75 150 75,clip,width=\textwidth]{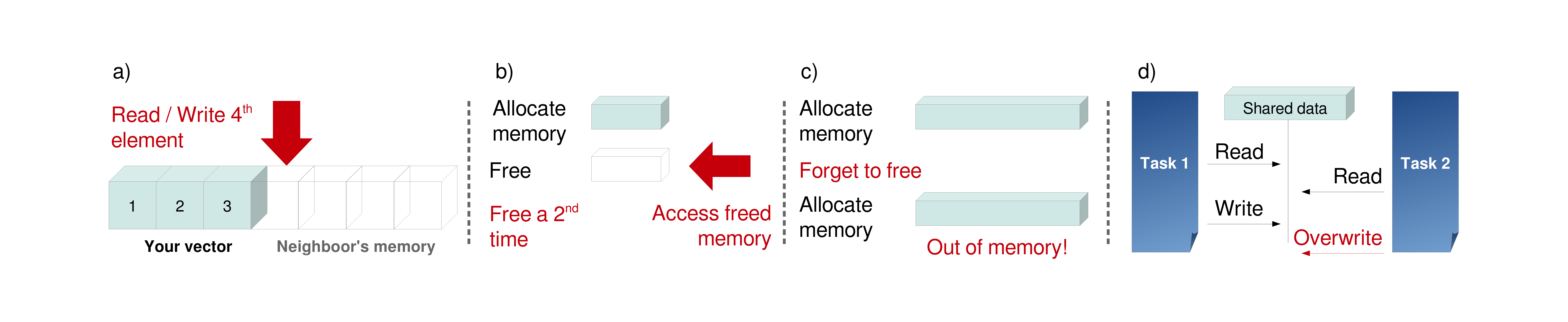} 
 \caption{Typical memory and concurrency problems with system programming languages such as Fortran, C or C++.}
   \label{fig1}
\end{center}
\end{figure}

Plenty of tools in astrophysics are developed using system programming languages such as Fortran, C or C++. 
These languages are known to provide high performance and fast executions but they rely heavily on the developer for concurrency and memory control, which may lead to common errors as shown in Fig.\,\ref{fig1}: a) access to invalid memory regions, b) dangling pointers and attempts to free already freed memory, c) memory leaks and, d) race conditions.
This can produce random behaviors and affect the scientific interpretation of the results.

The recently created language Rust prevents such problems and fields like bioinformatics \cite{Koster2015} have already started to take advantage of it.
Astroinformatics can benefit from it too.
We first discuss the general principles behind this new language and what makes it attractive when compared to more traditional languages such as C or Fortran. 
We then show that this language can reach the same performance as a Fortran N-Body simulator, Mercury-T\footnote{\href{http://www.emelinebolmont.com}{http://www.emelinebolmont.com}} \cite{Bolmont2015}, designed for the study of the tidal evolution of multi-planet systems.

\section{Rust}

Mozilla Research, motivated by the development of a new web browser engine \cite{Anderson2015}, released in 2015 the first stable version of a new open source programming language named Rust. 
It uses patterns coming from functional programming languages \cite{Poss2014} and it is designed not only for performance and concurrency, but also for safety. Rust introduces concepts like ownership, borrowing and variable lifetime, which:
\begin{enumerate}
\item[-] facilitates the automatic control of the lifetime of objects during compilation time. There is no need for manually freeing resources or for an automated garbage collector like in Java or Go;
\item[-] prevents the access to invalid memory regions;
\item[-] enforces thread-safety (race conditions cannot occur). 
These zero-cost abstraction features make Rust very attractive for the scientific community with high performance needs.
\end{enumerate}

\begin{figure}[htbp]
\begin{center}
 \includegraphics[trim=140 50 75 50,clip,width=\textwidth]{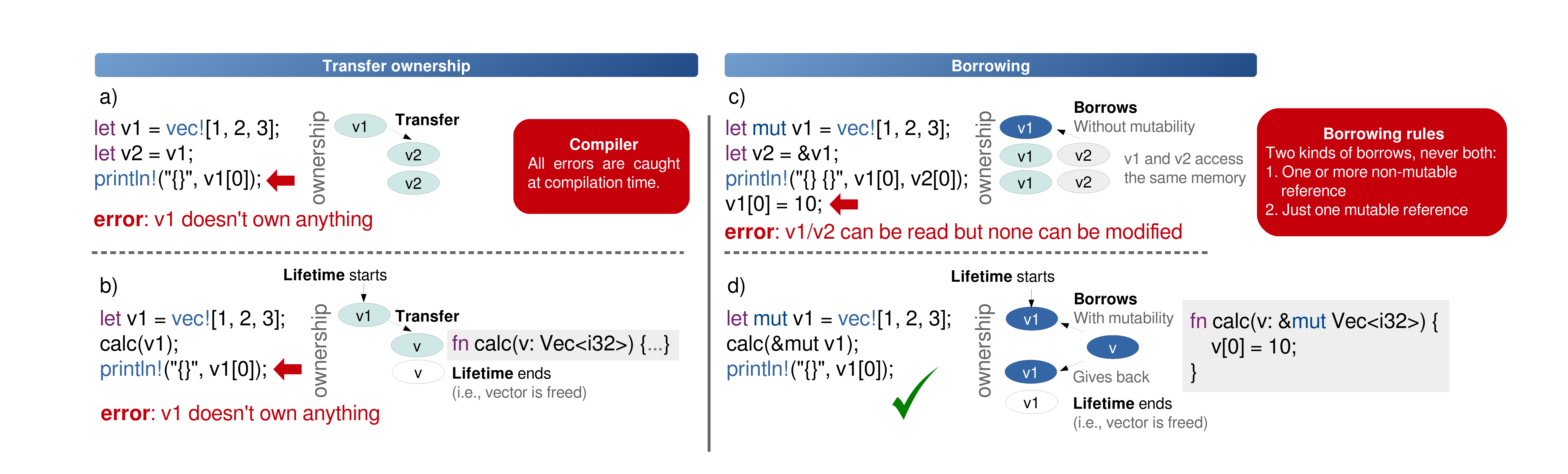} 
 \caption{Rust's ownership and borrowing concepts help us overcome the problems of traditional system programming languages.}
   \label{fig2}
\end{center}
\end{figure}

In Rust, variables are non-mutable by default (unless the mutable keyword is used) and they are bound to their content (i.e, they own it or they have ownership of it). When you assign one variable to another (case 'a' in Fig.\,\ref{fig2}), you are not copying the content but transferring the ownership\footnote{Except primitive types or types that implement the Copy trait.}, so that the previous variable does not have any content (like when we give a book to a friend, we stop having access to it). This transfer takes also place when we call functions (case 'b' in Fig.\,\ref{fig2}), and it is important to note that Rust will free the bound resource when the variable binding goes out of scope (at the end of the function call for case 'b'). Hence, we do not have to worry about freeing memory and the compiler will validate for us that we are not accessing a memory region that has already been freed (errors are caught at compilation time, before execution time).

Additionally, apart from transferring ownership, we can borrow the content of a variable (case 'c' in Fig.\,\ref{fig2}). In this case, two variables have the same content but none of them can be modified, thus protecting us from race conditions. Alternatively, we can borrow in a more traditional way (like when we borrow a book from a friend, he is expecting to get it back when we stop using it) like in case 'd' in Fig.\,\ref{fig2}, where the function borrows the content of a variable, operates with it (in this case, it could modify its content) and returns it to the original owner (not destroying it as shown in case 'b'). 

Exceptionally, all these rules can be violated if we make use of unsafe blocks, which is strongly discouraged but necessary in certain situation (e.g., dealing with external libraries written in Fortran, C or C++). If present, unsafe blocks allow us to clearly identify parts of the code which should be carefully audited, keeping it isolated and not making the whole program unsafe by default like in Fortran, C or C++.

\section{Assessment}

We explored the advantages and drawbacks of Rust for astrophysics by re-implementing the fundamental parts of Mercury-T \cite{Bolmont2015}, a Fortran code that simulates the dynamical and tidal evolution of multi-planet systems.

\medskip

We developed a simple N-Body dynamical simulator\footnote{\href{http://www.blancocuaresma.com/s/}{http://www.blancocuaresma.com/s/}} (without tidal effects) based on a leapfrog integrator in Rust, Fortran, C and Go (which provide a garbage collector for memory management). 
The software design and implementation does not include any language-specific optimization that a developer with basic knowledge would not do. 

We compiled the four implementations with an optimization level 3 (rustc/gfortran/gcc compiler) and the standard compilation arguments in the case of Go. 
We selected the best execution time out of five for an integration of 1 million years and the results are in Table~\ref{table}.
For this particular problem, Rust is as efficient as Fortran, and both surpass C and Go implementations

\begin{table*}[htbp]
\begin{center}
\caption{Best execution times of pure N-body simulations, for an integration time of 1 million years using a leap-frog integrator.}
\vspace{0.1cm}
\begin{threeparttable}
\begin{tabular}{c|c|c|c}
Rust & Fortran & C & Go  \\
\hline
0m13.660s & 0m14.640s & 2m32.910s$^{a}$ & 4m26.240s \\
\hline
\end{tabular} 
\label{table} 
\begin{tablenotes}
    \footnotesize
    \item[a] Time might be improved if language-specific optimizations were implemented.
\end{tablenotes}
\end{threeparttable}
\end{center}
\end{table*}

Based on Mercury-T, we implemented the additional acceleration produced by tidal forces between the star and its planets into our Rust and Fortran leapfrog integrators.

To test the codes, we ran a simulation of 100 million years with the same initial conditions as the case 3 described in the Mercury-T article \cite{Bolmont2015}, hence a single planet with a rotation period of 24 hours, orbiting a brown dwarf (0.08~$M_\odot$) at 0.018~AU with an eccentricity of 0.1.

\begin{figure}[htbp]
\begin{center}
 \includegraphics[width=8.5cm]{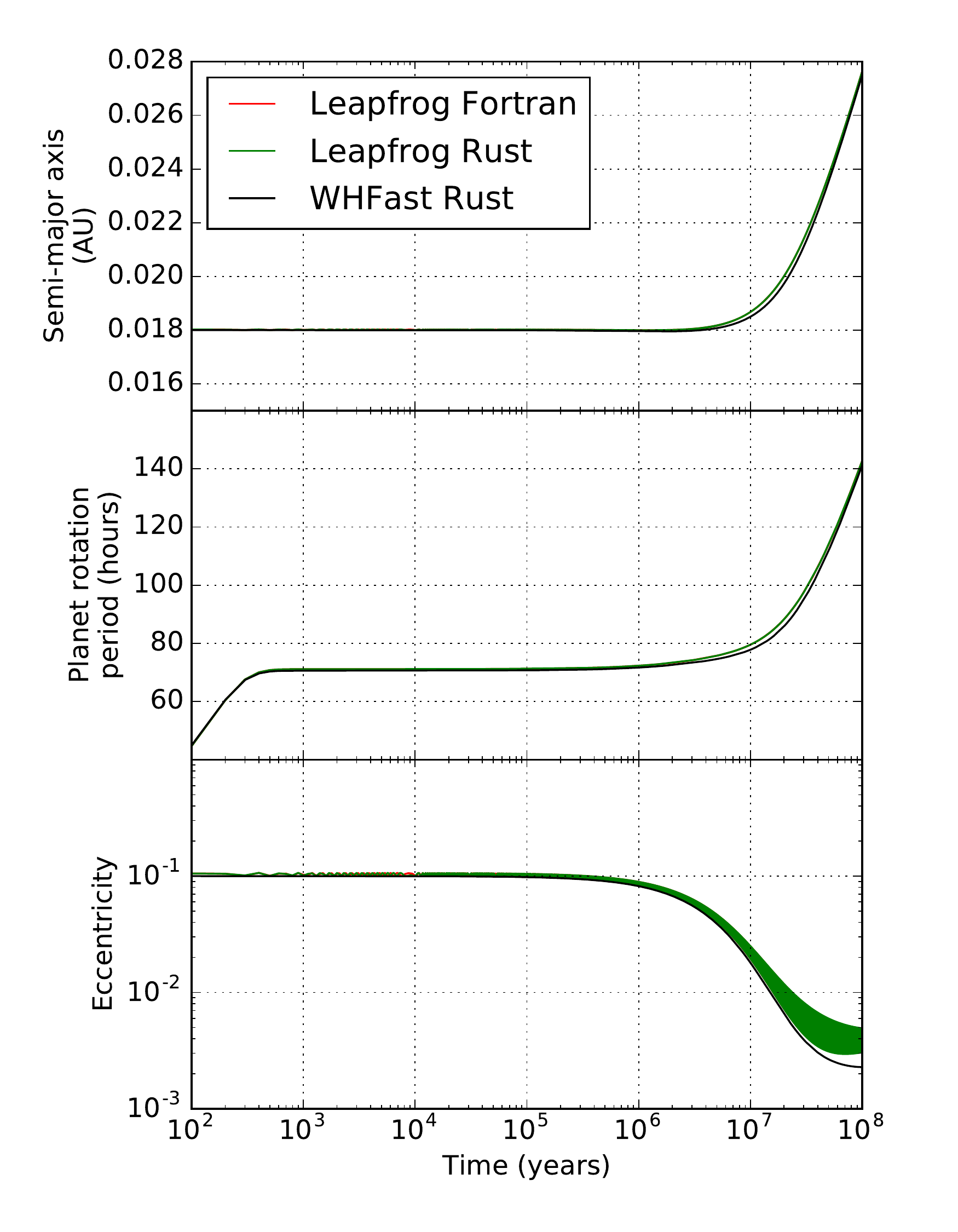} 
 \caption{Tidal evolution of a single planet orbiting a brown-dwarf with the three integrators. \textbf{Top}: evolution of the semi-major axis of the planet. \textbf{Middle}: evolution of the planet's rotation period. \textbf{Bottom}: evolution of the eccentricity of the planet.}
   \label{fig3}
\end{center}
\end{figure}

The results are shown in Fig.~\ref{fig3}, the Rust and Fortran code are practically identical and they reproduce a similar behavior to what is shown in the Mercury-T article.
Nevertheless, leapfrog is a very simple integrator and not very accurate. 
This can be seen in the eccentricity evolution, which is slightly different from the Mercury-T article and appears noisy. 
As an additional exercise, we implemented the WHFast integrator \cite{Rein2015} in Rust (black line in Fig.~\ref{fig3}).
This better integrator leads to a better agreement with Mercury-T thus demonstrating that a high level of accuracy can also be achieved with Rust.

\section{Conclusions}

We have shown the reliability of Rust as a programming language as opposed to Fortran, C or even Go.
Rust allows the user to avoid common mistakes such as the access to invalid memory regions and race conditions.
We have also shown that it is a competitive language in terms of speed and accuracy. 

The main challenge we experienced was the initial learning curve, it was necessary to really understand and get used to the ownership and borrowing concepts. 
Once the paradigm shift is done, the benefits are immediate.
We therefore encourage the community to consider Rust as a language that will help us produce good quality, memory safe, concurrent and high-performance scientific code.





\end{document}